\documentclass[aps,prb,twocolumn,superscriptaddress,showpacs]{revtex4}
\usepackage{graphicx}
\usepackage{calc}
\usepackage{bm}

\begin{document}

\title{Thermoelectric properties of the interacting two dimensional electron gas in the diffusion regime.}

\author{V.~T. Dolgopolov}
\affiliation{Institute of Solid State Physics RAD, Chernogolovka,
Moscow District, 142432, Russia}

\author{A. Gold}
\affiliation{Centre d'Elaboration de Materiaux et d'Etudes Structurales (CEMES/CNRS) and UniversitŽ Paul Sabatier, 29 Rue Jeanne Marvig, 31055 Toulouse, France}

\date{\today}

\begin{abstract}
We demonstrate that kinetic coefficients related to thermoelectric
properties of the two dimensional electron gas in the diffusive
regime are strongly influenced  by electron-electron interaction.
As an example we consider the thermoelectric coefficients of the
diluted two-dimensional electron gas in Si(100) MOSFET's in the
presence of charged-impurity scattering.  We  find that the screening
anomaly at $q=2k_F$, also responsible for Friedel oscillations,
leads at low electron densities to a large change in
the thermoelectric coefficient for the thermopower.
\end{abstract}


\maketitle

Since a long time thermoelectric effects in  metals are considered
in textbooks, see for instance the book by Ashcroft and
Mermin.\cite {AM} Nevertheless, all known results of explicit
calculations for two dimensional metals are made within the
random-phase approximation and are, therefore, only valid for weak
electron-electron interaction, which means for very high electron
densities.\cite {pines} In a number of publications
\cite{karavolas,butcher,fletcher1,hwang,cm} the attempt was made
to describe real experiments and these results were extrapolated
down to low electron density where the Coulomb interaction is
strong. We argue in this paper that an important term for the
calculation of thermoelectric coefficients was neglected in this
extrapolation.

The present calculation  demonstrates that the interaction leads
to a strong change in the kinetic characteristics of the two
dimensional electron gas, such that all textbook predictions for the
Seebeck and the Peltier effect became invalid in the limit of low
electron densities. From experiment \cite{shashkapust,
shashanis,Klimov} and the theory \cite{senatore,sarma,khodel}
it is well known that electron-electron interaction is the origin of
a significant  change in the electron dispersion law. The influence of
interaction on kinetic coefficients is an additional effect which
is not related to a change of the dispersion law of electrons.

The presence of temperature $T$ and electric potential gradients
(electric field ${\bf E}$) in the metallic system is accompanied
by the flow of an electric current ${\bf I}$ and  a heat current
${\bf J}$:

\begin{eqnarray} {\bf I}=L^{11}{\bf E} + L^{12} (-\nabla T), \\
 \nonumber
   {\bf J} =L^{21}{\bf E} +L^{22}(-\nabla T), \nonumber
   \\  L^{21} = TL^{12} .\nonumber \label{00}
\end{eqnarray}

In the heat transfer phonons prevail at high temperatures and
thermoelectric effects are caused by the electron-phonon interaction.
In the low temperature limit electrons give the main contribution
to the heat flow and the thermoelectric effects at low
temperatures are caused by the diffusion of electrons.

Below we discuss thermoelectric effects due to diffusion in the
two dimensional electron gas in the presence of electron-electron
interaction. At low electron densities $n$ the interaction effects
become very large. In the best two dimensional electron systems
\cite{shashkinUFN} the value of dimensionless Wigner-Seitz
parameter $r_s=1/(\pi n a_B^2)^{1/2}$, describing the importance of
the interaction effects, is about 10. Here $a_B$ is the effective Bohr radius.

Parameter $r_s=10$ means that the potential energy exceeds the
Fermi energy (kinetic energy) by about one order of magnitude.
Recently very clear expressed evidences of the strong
electron-electron interaction were experimentally observed in
Si-MOSFETs (see e.g. Refs. \cite{shashkapust, shashanis,Klimov}).
Having this fact in mind we will apply our consideration to (100)
Si-MOSFET with high mobility. Of course our results will be valid
for other two dimensional electron and hole system.

To the best of our knowledge thermopower calculations in the low
density range with strong electron-electron interaction were never
realized, despite the fact that the experimental activity in
two-dimensional electron systems is directed in lowering the
electron density. We restrict our consideration to the region of
low electron densities, where  the system is still chacterized by
a metallic conductivity $\sigma>>e^2/h$. We  consider a simple
model of a two-dimensional electron system  in the metallic regime
and we use the relaxation-time approximation. A gas of Landau
quasiparticles has, renormalized by interaction effects, a mass
$m$ and a thermodynamic density of states $\rho_F$. \cite{gold11}
 In our calculation we neglect with heat transfer by low
energy collective modes \cite{aleiner2} which has minor influence
on the thermoelectric coefficients. Having in mind a low electron
density system we suppose charged-impurity scattering by
impurities randomly distributed at the $Si/SiO_2$ interface with
an impurity concentration $n_i$. The extension effects of the
electron wave function into the silicon are described by the
formfactor $F_i(q)$ for the electron-impurity interaction and by
$F(q)$ for the electron-electron interaction.

The appearance of thermoelectric effects is caused by the fact
that the diffusion coefficient near the  Fermi level is a function
of the energy. A temperature gradient produces two diffusion
streams of electrons: "cold" electrons moves along the temperature
gradient and "hot" ones in the opposite direction. If the
diffusion coefficient is independent of energy, charge transfer is
absent. In the case of an energy independent transport scattering
time the diffusion coefficient increases with energy, which leads
to a charge flow opposite to the temperature gradient, i.e.,  to
the appearance of a charge flow typical for  ideal electron
system. The contribution from an energy dependent transport
scattering time should be added to this term.

For a strong electron-electron interaction the screening function
and the transport scattering time show a peculiar behavior at the
Fermi energy. Scattering processes for electrons below and above
the Fermi level are different due to the difference in the
screening properties \cite{gold1,gold2} which can be also
interpreted as additional scattering due to Friedel oscillations
for high energy electrons \cite{Zala}. On the first glance this
peculiar behavior of the screening function gives only small
contributions to the transport scattering time and can be omitted
for zero temperature.\cite{gold1,gold2} However, the contribution
to the energy dependence of the diffusion coefficient has the same
order of magnitude as other terms and is therefore very important.
This effect was never considered in the past and will be shown in
the following.

Below we show that the "ideal electron system" contribution to the
energy dependence of the diffusion coefficient for a strong
electron-electron interaction is compensated in part by the energy
dependence of the transport scattering time. As a result the
Seebeck effect and Peltier effect at low electron densities are
expected to be significantly smaller than those for a non-
interacting electron gas and even might change the sign.
\cite{rem}

Thermoelectric coefficients are sensitive to the energy dependence
of the transport scattering time and are equal to
\begin{equation}
L^{12}_d=\frac{L^{21}_d}{T}=\frac{g_v e}{2 \pi^2}\int\int{\textbf
v_k} {\textbf v_k} \tau_0(\varepsilon)
\frac{(\varepsilon-\varepsilon_F)}{T} (\frac{dS}{|v_k|})(-
\frac{\partial f^0}{\partial\varepsilon})d\varepsilon \label{8}
\end{equation}
where $v_k$ is the electron velocity, $dS =|k|d\phi$, $f^0$ is the Fermi-Dirac
distribution function, and $g_v$ is the valley degeneracy.

The transport scattering time $\tau_0(\varepsilon)$ for a particle of energy
$\varepsilon$ in lowest order of the random potential $<|U(q)|^2>$ is given by
\cite{ando,gold1}
\begin{equation}
\frac{1}{\tau_0(\varepsilon)}=\frac{1}{2\pi\hbar\varepsilon}
\int_{\text{0}}^{\text{2k}} dq \frac{q^2}{(4k^2-q^2)^{1/2}}
\frac{<|U(q)|^2>}{\epsilon(q)^2},
\end{equation}
where $<|U(q)|^2>=n_iV(q)^2 F_i(q)^2$ with $V(q)=2\pi e^2/
\epsilon_Lq$, and
$\epsilon_L=\frac{1}{2}(\epsilon_{sc}+\epsilon_{ins})$ is the
dielectric constant of the background material.\cite{ando}

For a non-interacting electron gas with short-range random
potential as disorder the time $\tau_0$ is independent of energy
and only the "ideal electron system" contribution to the
thermoelectric coefficients exists. In this case the thermopower
is given by the well known expression

\begin{equation}
S_{d0} =-\frac{L_{12}}{L_{11}}=- \frac{L_{12}}{\sigma_0}=- \frac{\pi g_v k_B^2 m T}{3e n},     \\
k_B T \ll \varepsilon_{F}.
 \label{1b}
\end{equation}
Here $m$ is electron effective mass.

 With the aim to take into
account the screening of impurities and especially the  anomalous
screening for $q>2k_F$ we write
\begin{eqnarray}
\epsilon(q) = \epsilon_{1}(q)[1-\frac{V(q)(1-G(q))\rho_F
F(q)}{\epsilon_{1}(q)}\nonumber
\\ (1-\frac{4k_{F}^2}{q^2})^{1/2} \Theta(q^2-4k_{F}^2)], \label{tau}
\end{eqnarray}
where $\epsilon(q)$ is the screeing function of the two
dimensional electron system, $G(q)$ describes the local-field
correction for exchange and correlations effects not taken into
account in the random-phase approximation.\cite{singwi}
$\Theta(x)$ is the step function with $\Theta(x)=1$ for $x\geq0$
and $\Theta(x)= 0$ for $x<0$. $\epsilon_{1}(q)$ is defined by
\begin{equation}
\epsilon_{1}(q) = 1+V(q)F(q)(1-G(q))\rho_F  \label{epsilon}
\end{equation}

We use for the explicit calculation the local-field corrections in
the Hubbard (H) approximation $G(q)=G_H(q)$ where an analytical
expression is available as
\begin{equation}
 G_H(q)
=\frac{1}{2g_v}\frac{q}{(q^2+k^2_F)^{1/2}}   . \label{H}
\end{equation}

For Si(100) MOSFET
structures the valley degeneracy factor is $g_v=2$.

With Eqs.~(\ref{tau},\ref{epsilon}) we can write inverse
transport scattering time as
\begin{equation}
\frac{1}{\tau_0(\varepsilon)}=
\frac{1}{\tau_0(\varepsilon_F)}+\frac{1}{\tau_1(\varepsilon)}.
\label{3}
\end{equation}

For low temperatures only terms proportional to
$(\varepsilon-\varepsilon_F)/\varepsilon_F$ need to be taken into
account in the energy development of the transport scattering time
in Eq.~\ref{3}. Only these terms will be important for the
calculation of $L_{ij}$ for low temperatures.

We evaluate the energy independent part as
\begin{eqnarray}
\frac{1}{\tau_0(\varepsilon_F)}\simeq\frac {4m}{\pi\hbar^3}
\frac{<|U(2k_F)|^2>}{[\epsilon_1 (2k_F)]^2}
\int_{\text 0}^{\text 1} \frac{x^2 dx}{[1-x^2]^{1/2}}=  \nonumber   \\
=\frac{m}{\hbar^3} \frac{<|U(2k_F)|^2>}{[\epsilon_1 (2k_F)]^2} .
\label{4}
 \end{eqnarray}

The contributions to $\frac{1}{\tau_1(\varepsilon)}$ consist of two
terms:
\begin{eqnarray}
\frac{1}{\tau_1(\varepsilon)} \simeq
(\varepsilon-\varepsilon_F)\frac{m}{\hbar^3}
\frac{\partial}{\partial\varepsilon}[\frac
{<|U(2k)|^2>}{[\epsilon_1 (2k)]^2}]_{ \varepsilon=\varepsilon_F} +
 \frac{(\varepsilon-\varepsilon_F)}{\varepsilon_F}\frac{2m}{\hbar^3}  \nonumber \\
 \Theta(\varepsilon-\varepsilon_F) V(2k_F)F(2k_F)(1-G(2k_F))\rho_F \frac{<|U(2k_F)|^2>}{\epsilon_1
 ^3 (2k_F)}.  \nonumber  \\
 \label{5}
\end{eqnarray}
As we will see, the first is the regular term of the expansion and gives only a small
contribution to thermoelectric coefficients at low electron
densities. More important is the second non-analytic term in the energy expansion of the transport scattering time.

From Eq.~(\ref{4}) and Eq.~(\ref{5}) it follows
\begin{equation}
\frac{1}{\tau_1(\varepsilon)} = \frac{1}{\tau_0(\varepsilon_F)}
\frac{\varepsilon-\varepsilon_F}{\varepsilon_F}[B(\varepsilon_F)+2C(\varepsilon_F)\Theta(\varepsilon-\varepsilon_F)],
\label{7}
\end{equation}
where $B(\varepsilon_F)=[B_1(\varepsilon_F)+B_2]$ is the sum of
two terms. One of the terms in square brackets describes the
contribution of form factors and is equal to
\begin{equation}
B_1(\varepsilon_F)=\frac{1}{8}(15-33\frac{\epsilon_{ins}}{\epsilon_{sc}})\frac{k_F}{b}-\frac{1}{\epsilon_1(2k_F)},
\label{B1}
\end{equation}
where $b=(48\pi m_\bot e^2N^*/\epsilon_{sc})^{1/3}$,
$N^*=N_{depl}+\frac{11}{32} n$ , and  $\frac{k_F}{b}\ll 1$.

The second term $B_2$ is caused by the local-field correction and is
independent of electron density and equal to
\begin{equation}
B_2 \simeq\frac{1}{5^{3/2} g_v-5}  . \label{B}
\end{equation}
For $g_v=2$ we have $B_2\simeq 0.06$ and for  $g_v=1$ we have
$B_2\simeq 0.16$. The function $C(\varepsilon_F)$ is expressed by:
\begin{equation}
C(\varepsilon_F)\simeq1-\frac{1}{\epsilon_1(2k_F)}. \label{C}
\end{equation}
This $C(\varepsilon_F)$ term describes the non-analytical behavior
of the energy expansion of the transport scattering time and will
be the origin of anomalous temperature dependencies. The transport
scattering time is then given by
 \begin{equation}
\tau(\varepsilon)=
\tau_0(\varepsilon_F)(1-\frac{\varepsilon-\varepsilon_F}{\varepsilon_F}[B(\varepsilon_F)+2C(\varepsilon_F)\Theta(\varepsilon-\varepsilon_F)])
\label{7a}
\end{equation}

After substitution of  Eq.~(\ref{7a}) into Eq.~(\ref{8}) we
obtain the final result for $g_v=2$

   \begin{eqnarray}
   L_d^{12} \simeq [- 0.06
   +\frac{2}{\epsilon_1(2k_F)}-
    \frac{15-33\frac{\epsilon_{ins}}{\epsilon_{sc}}}{8}\frac{k_F}{b}] \frac{\pi e \tau_0(\varepsilon_F)}{3} k_B^2
    T. \nonumber \\
   \label{9}
   \end{eqnarray}
Note that the "ideal electron system" contribution and the main
term in the non-analytical energy dependence of the transport
scattering time cancel each other.
  In the limit of very small electron density  Eq.~(\ref{9}) reads
\begin{eqnarray}
L_d^{12} \simeq [-0.06+ \frac{2}{g_v
r_s\frac{m}{m_b}(\frac{g_v}{2})^{\frac{1}{2}}
(1-\frac{0.45}{g_v})}- \nonumber \\
0.47(\frac{k_F}{b})]\frac{\pi e \tau_0(\varepsilon_F)}{3} k_B^2 T.
\label{10a}
\end{eqnarray}
Here effective mass $m$ is  renormalized by interaction effects
and $m_b$ is the value for the band mass.

Eqs.~(\ref{9},\ref{10a}) show us that within the accuracy of our
calculations the diffusion thermoelectric coefficients in the
regime of strong electron-electron interaction are significantly
lower than the values expected for the case of a normal metal with
weak interaction. Similar behavior is observed in experiment.
~\cite{fletcher} Moreover, there is a contribution which leads to
the sign change  of the thermoelectric coefficient $L_d^{12}$ at
very low electron densities, reachable in remote doped Si/SiGe
heterostructures, where the Coulomb interaction is strong.

In our model the thermopower at low temperatures is given by the
following expression

\begin{equation}
S_d =-\frac{\pi^2 k_B^2 T}{3e\varepsilon_F}[1-B(\varepsilon_F)
-C(\varepsilon_F)],     \\
k_B T \ll \varepsilon_{F}.
 \label{11}
\end{equation}
It clearly shows the importance of the Coulomb interaction by the
factor $C(\varepsilon_{F})$ containing explicitly and in
analytical form the interaction potential including exchange and
correlation. The appearance of this factor $C(\varepsilon_{F})$ in
the formula for the thermoelectric power of an interacting 2DEG is
the new result of our paper.
This factor does not appear in three-dimensional systems because
there the non-analytical term in the scattering time is much
smaller. But we expect that a similar term should exist in the
one-dimensional interacting electron gas.

 The parameter
$B(\varepsilon_F)$ also can be written as
\begin{equation}
B(\varepsilon_F)=-\frac{\varepsilon_F}{\tau_0(\varepsilon_F)}[\frac{d\tau_0(\varepsilon)}{d\varepsilon}]_{\varepsilon=\varepsilon_F^-}
.
 \label{AG}
\end{equation}
We stress that the derivative must be taken at an energy from
below the Fermi energy $\varepsilon=\varepsilon_F^-$ in order to
avoid the non-analytic behavior of the transport scattering time.
This term $B(\varepsilon_F)$ is in agreement with earlier results
on the thermopower. \cite{karavolas,butcher,fletcher1,cm}

For (100)Si MOSFET systems with intermediate electron density
$(3-10)*10^{11}cm^{-2}$ we evaluate the thermopower as
\begin {equation}
S_d \simeq - \frac{r_s^2a_B^2 m}{3e\hbar^2} k_B^2T[1.3
\frac{m_b}{mr_s}-0.47\frac{k_F}{b}-0.06] . \label{SS}
\end{equation}
It is interesting to compare Eq.\ref{SS} and  Eq.\ref{1b}.
At low density ($r_s \gg1$) the absolute value of thermopower is
strongly decreased and the dependence on the electron density is
expected to be rather weak, because $S_d \propto r_s $.

Until now we have discussed only the low temperature behavior of
the thermoelectric coefficients. In experiment the temperature
interval for the diffusion regime conventionally is defined as the
interval in which the thermopower is proportional to the
temperature. For the thermopower in the diffusive regime we get
linear corrections, which are anomalous due to the anomalous
screening in two dimensions. We conclude that $S_d/T=c_1+c_2T$.
$c_1$ was calculated in the present paper.

For more realistic calculations and the comparison with real
experimental results one needs all disorder parameters such as the
density of impurities, the parameters for the interface-roughness
scattering and a realistic form of confinement potential
(depletion density).\cite{zianni} Moreover, for the comparison
between theory and experiment it would be useful to measure for
the same sample as function of density and temperature the
conductivity and the thermopower.

The main message of our calculation consist in the result that the
anomalous screening in two dimensions has a much larger impact for
the Seebeck and the Peltier effect than on the conductivity. The
effect is such large that at low carrier density all calculations,
not taking into account this contribution, are unable to predict
correctly even the order of magnitude for the Seebeck and the
Peltier effect. The corresponding contributions to the
conductivity [$L_{11}$ in Eq.(\ref{00})] and the thermoconductivity
[$L_{22}$ in Eq.(\ref{00})] are only small corrections, linear in
temperature,\cite{gold1,gold2} and disappear at very low temperature.

Our theory generalizes earlier theoretical results obtained for
the diffusion thermoelectric coefficients in the interacting
two-dimensional electron gas. The additional contribution we
obtain depends strongly on the strength of the Coulomb interaction
including exchange-correlation effects via the local-field
correction.

We gratefully acknowledge discussions with A.~A. Shashkin and
V.~F. Gantmakher. This work was supported by the RFBR, the Russian
Ministry of Sciences, and by the Programmes of RAS.


\begin{thebibliography}{apssamp}
\bibitem{rem} We find a similar change of thermoelectric coefficients
in the model of Ref.~\cite{Zala}, but it is
impossible to evaluate the accuracy of the received result because the
non-singular part of the transport scattering time cannot be calculated within this
model.
\end{thebibliography}
\end{document}